\documentclass[review]{elsarticle}

\pdfoutput=1
\usepackage{amssymb}
\usepackage{amsmath}
\usepackage{float}
\usepackage{morefloats}
\usepackage{dblfloatfix}
\usepackage{graphicx}
\usepackage{xcolor}
\usepackage[ruled, boxed, vlined]{algorithm2e}
\usepackage{algorithmic}
\usepackage{multirow}
\usepackage[caption=false,font=footnotesize]{subfig}
\usepackage{url}

\makeatletter
\def\ps@pprintTitle{%
 \let\@oddhead\@empty
 \let\@evenhead\@empty
 \def\@oddfoot{\centerline{\thepage}}%
 \let\@evenfoot\@oddfoot}
\makeatother

\begin{document}

\begin{frontmatter}

\title{Analysis of Deformation Fields in Spatio-temporal CBCT images of lungs for radiotherapy patients}

\author{Bijju~Kranthi~Veduruparthi\corref{corra}}
\ead{bijjuair@gmail.com}
\author{Jayanta~Mukherjee\corref{}}
\author{Partha~Pratim~Das\corref{}}
\address{Department of Computer Science and Engineering, Indian Institute of Technology, Kharagpur}



\author{Mandira~Saha\corref{}}
\author{Raj~Kumar~Shrimali\corref{}}
\author{Sanjoy~Chatterjee\corref{}}
\address{Department of Radiation Oncology, Tata Medical Center, Kolkata}

\author{Soumendranath~Ray\corref{}}
\address{Department of Nuclear Medicine, Tata Medical Center, Kolkata}

\author{Sriram~Prasath\corref{}}
\address{Department of Medical Physics, Tata Medical Center, Kolkata}

\cortext[corra]{Corresponding author}

\begin{abstract}
Deformable registration of spatiotemporal Cone-Beam Computed Tomography (CBCT) images taken sequentially during the radiation treatment course yields a deformation field for a pair of images.
The Jacobian of this field at any voxel provides a measure of the expansion or contraction of a unit volume.
We analyze the Jacobian at different sections of the tumor volumes obtained from delineation done by radiation oncologists for lung cancer patients.
The delineations across the temporal sequence are compared post registration to compute tumor areas namely, unchanged ($U$), newly grown ($G$), and reduced ($R$) that have undergone changes.
These three regions of the tumor are considered for statistical analysis. In addition, statistics of non-tumor ($N$) regions are taken into consideration.
Sequential CBCT images of 29 patients were used in studying the distribution of Jacobian in these four different regions, along with a test set of 16 patients.
Statistical tests performed over the dataset consisting of first three weeks of treatment suggest that, means of the Jacobian in the regions follow a particular order. 
Although, this observation is apparent when applied to the distribution over the whole population, it is found that the ordering deviates for many individual cases.
We propose a hypothesis to classify patients who have had partial response (PR). Early prediction of the response was studied using only three weeks of data.
The early prediction of response of treatment was supported by a Fisher's test with odds ratio of 5.13 and a p-value of 0.043.
\end{abstract}


\end{frontmatter}


\section{Introduction}
Lung cancer patients undergo a six week fractionated course of radiotherapy as part of  their curative treatment. Radiation therapy planning requires identifying the tumor based on a pre-treatment planning computed tomography(CT) taken in treatment position followed by localization of the tumor by radiation oncologists. During the process nearby “organs at risk” are also identified to ensure that radiation induced damage is avoided. This localized volume is then planned to receive a specified (often 60Gy) dose of radiation in thirty fractions over a period of six weeks. Modern radiation therapy machines can acquire a cone beam CT scan(CBCT) image on a daily or more commonly, a weekly basis prior to delivery of the radiation to negate any positional changes in the identified tumor. Such image guided radiation therapy is often the standard of care for the curative therapy of Lung cancers. Adaptive Radiation Therapy (ART) requires oncologists to adapt the radiation volume based on spatio-temporal changes in the tumor, and has the potential to personalize radiation therapy based on such changes. Registration of such images therefore remain a key area of concern and could not only help identify the target accurately, but could also potentially prognosticate the outcomes in such patients.

From a big dataset of radically treated lung cancer patients in a tertiary care center, we randomly selected their pre-treatment CBCT and on-treatment serial CBCT images. Weekly CBCTs of the patients are delineated by experienced radiologists for approval, and treatment. Each of the CBCTs are first delineated by an expert radiologist, followed by correction and approval of a second senior radiologist. Finally, the approved delineations are used by the radiation planning team to position the patient for radiotherapy.

The work studies the statistical analysis of deformation fields obtained by registering pre-treatment CBCT and serial on-treatment CBCT images of lung cancer patients who underwent radiotherapy. Using only the first three weeks of treatment as data, we aim to predict the response of the patient to treatment before the therapy completes. To ensure clinical relevance, we used the actual target volumes delineated by the clinical oncologist on pre-treatment CBCT for treatment and thereafter on CBCT to assess the changes in the targets. The delineations were used to categorize the tumor into different regions. We study the Jacobian statistic on the deformation fields in these regions and study how it varies across patients. We observe that patients satisfying the proposed hypothesis have better response to treatment than those who do not satisfy the hypothesis. Using this hypothesis, we are able to classify patients into PR and non-PR categories, and support the hypothesis using Fisher's association test.

\subsection{Related work}

Computational Anatomy \cite{Miller2009} is a study of anatomy using mathematical and computational models. The deformable template that exhibits diffeomorphism, a bijection between anatomical coordinates is sometimes used in such a study. In \cite{Miller2004}, landmarks around the area of interest were used to track changes, where the Jacobian maps showed the direction of growth in ventral and dorsal parts of the hippocampus of a mouse. Deformation fields obtained after deformable image registration were earlier used in \cite{Amelon2010} for characterizing different regions of the lung, based on the motion properties like directional change, volume change and nature of change. In another study \cite{cao2012tracking}, the deformation fields were used to track changes in tissue volumes. Different regions of the lung were segregated into blocks, and average Jacobian in these regions was observed to report lung activity and for tracking change in volume of tissues.
In \cite{lorenzi2013sparse}, anatomical changes in the brain were observed to be different across spatial scales using the Helmholtz decomposition of the deformation field. A difference of Gaussians (DOG) operator was applied on the irrotational component of the decomposition to identify the areas of maximal volume change. However, this method was applied only to brain MR images. 
In another work \cite{lepore2007mean}, an average template was generated using registration of brain MR images. Statistical analysis on the deformation tensors with respect to this average template was performed to find anomalies in brain. Longitudinal analysis of spatiotemporal data to measure the growth of hippocampus in the brain is reported in \cite{durrleman2013toward}. Here, differences between growth trajectories are used to estimate a mean growth behavior of a population. This is compared with subjects to identify delay in growth for children diagnosed with autism.
Biological growth using deformation fields is mathematical modeled in \cite{grenander2007pattern}. The model is called a Growth by Random Iterated Diffeomorphisms (GRID). This model uses random seeds with radial deformations around it, to capture growth based deformations. In \cite{portman2009direct} and \cite{srivastava2005maximum}, the growth variables in GRID model are directly captured from image data. All the aforementioned approaches use Jacobian for measuring growth and decay from images using nonlinear registration. 

\subsection{Contribution}
This work presents a statistical analysis of deformation fields obtained by registering CBCT images of lung cancer patients who underwent radiotherapy. In our work, we also make use of delineation of tumor in CBCT images by radiologists. This is the first work of its kind on a dataset annotated by radiologists, and studying the behavior across population. The delineations were used to categorize the tumor into different regions. We study the Jacobian statistic on the deformation fields in these regions, and study how it varies across patients. Patients satisfying our proposed hypothesis were found to have better response to treatment. The results of the classification suggest that early prediction is feasible.

\section{Method}\label{method}
\subsection{Image Registration}\label{img_reg}
Given two images $S$, $T \in \mathbb{R}^3$, the goal of image registration is to find a transformation
$\vec{g} : \mathbb{R}^3 \mapsto \mathbb{R}^3$ that maps/aligns $S$ onto $T$.
The approach of computing both the forward (registering $S$ to $T$) and reverse (registering $T$ to $S$) transformations is termed as bidirectional symmetric registration. This process involves computation of inverse of the deformation field.
A symmetric log-domain based nonlinear image registration technique was proposed in \cite{vercauteren2008symmetric}. The deformation field obtained using this method is diffeomorphic and the true inverse can be computed efficiently at a very low cost. Hence this method is suitable for computational anatomy. This technique, however, is based on the sum of squared differences (SSD) in the intensities of the image. This technique is not quite applicable in our scenario, on account of the high amount of noise present in CBCT images. The noise in CBCT images is due to several artifacts arising from usage of low energy beams to produce the image. Further improvement in the results of the registration technique is reported  in \cite{Lorenzi2013}, where the symmetric Local Correlation Coefficient (LCC) provided a robust similarity measure. The technique was found to give smooth deformation fields owing to the regularization imposed on the total energy. We use this technique to perform non-linear image registration.

The deformation field that warps $S$ to $T$ is
denoted as $\vec{\phi} = \vec{z} - \vec{g}(\vec{z})$, where $\vec{z}$ is the set of points in $S$ that are mapped to corresponding points $\vec{z} - \vec{g}(\vec{z})$ in $T$ with displacement field $\vec{g}(\vec{z})$. Such a deformation $\vec{\phi}$ yields proper alignment of the two images.
Trilinear interpolation was used to compute the warping of deformation field at non-integer coordinates.

The deformation field obtained using this method is diffeomorphic and the inverse can be computed efficiently at a very low cost. Moreover, this method guarantees invertibility of the deformation and is invertible. Hence this method is suitable for computational anatomy.

\subsection{Analysis of deformation fields}\label{def_study}
The deformation index for interpreting the information in a deformation field is the determinant of Jacobian, commonly referred in the literature as simply Jacobian.
The Jacobian of a deformation field $\phi$ is defined as:
\begin{equation}
\label{eq:JacobianDet}
J(\vec{\phi}(\vec{z})) = \begin{bmatrix} \frac{\partial \phi_1(\vec{z})}{\partial z_1} & \frac{\partial \phi_1(\vec{z})}{\partial z_2}
  & \frac{\partial \phi_1(\vec{z})}{\partial z_3} 
  & \\ \frac{\partial \phi_2(\vec{z})}{\partial z_1} & \frac{\partial \phi_2(\vec{z})}{\partial z_2} & \frac{\partial \phi_2(\vec{z})}{\partial z_3} 
  & \\ \frac{\partial \phi_3(\vec{z})}{\partial z_1} & \frac{\partial \phi_3(\vec{z})}{\partial z_2} & \frac{\partial \phi_3(\vec{z})}{\partial z_3}
\end{bmatrix}
\end{equation}

\begin{figure*}[!ht]
\centering
\subfloat[]{\includegraphics[width=0.59\textwidth]{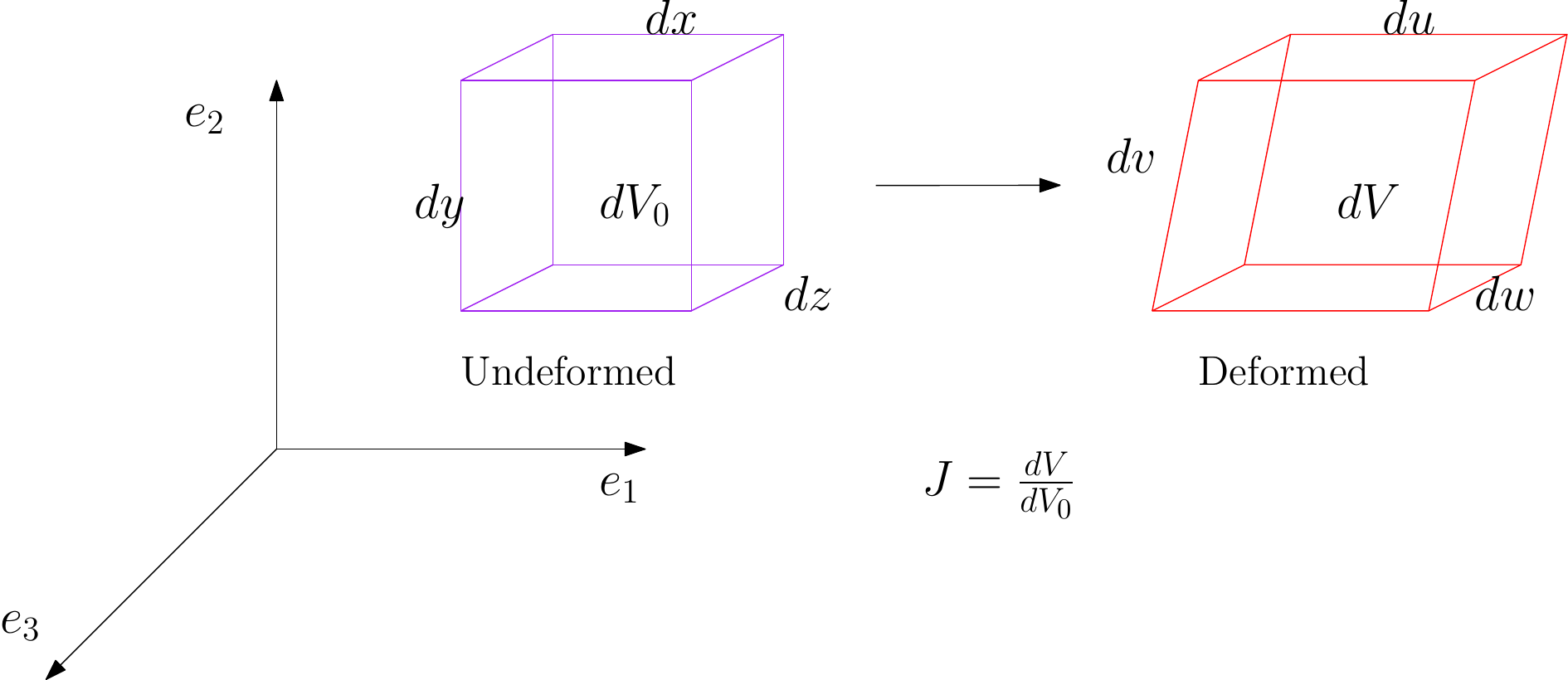}
\label{fig:JacobianVolume}}
\subfloat[]{\includegraphics[width=0.39\textwidth]{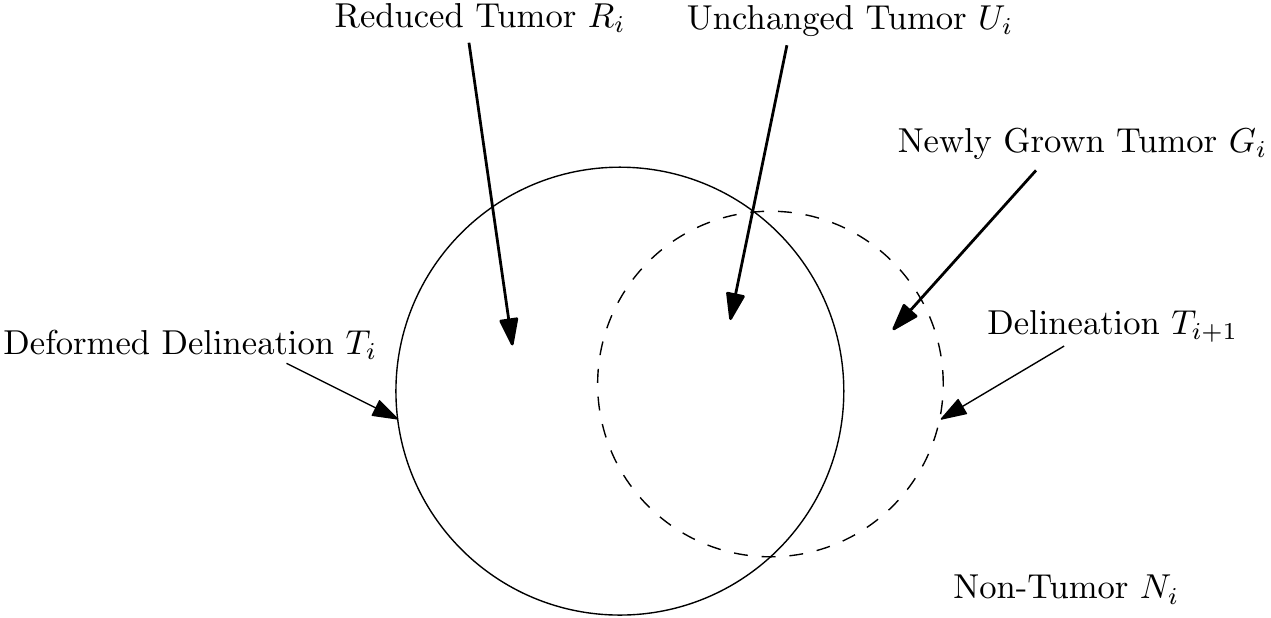}
\label{fig:GrowthDecayModel}}
\caption{(a) Jacobian computes the ratio of the deformed volume to the reference undeformed volume. (b) Growth-Decay model showing different regions of the tumor for two overlapping synthetic tumors. Solid line shows source tumor delineated volume. Dotted line shows the target volume.}
\label{fig:JacobianModel}
\end{figure*}

The determinant of $J(\vec{\phi}(\vec{z}))$, denoted as $|J(\vec{\phi}(\vec{z}))|$ or in short $J$ , can also be expressed in terms of its eigen values. The Jacobian $(J)$ measures the local volume change with respect to that of a unit cube.
This can be visualized in Fig.~\ref{fig:JacobianVolume}.
$J$ varies between $0$ and $\infty$. When $J = 1$, it means there is no change in volume.
A value of $J < 1$ denotes, net contraction; and $J > 1$ denotes, net expansion.

We computed the Jacobian in different areas of the 3D image and calculated the variation in the Jacobian statistic. The categorization can be visualized in Fig.~\ref{fig:GrowthDecayModel}. The Jacobian is computed individually in the following regions:
\begin{itemize}
\item \textbf{Deformed Tumor Region} in week $i$ ($T_i$): Set of tumor voxels when the image is deformed.
\item \textbf{Tumor Region} in week $i$ ($T^{\prime}_i$): Set of tumor voxels when the image is undeformed.
\item \textbf{Deformed Non-Tumor Region} in week $i$ ($N_i$): Set of non-tumor voxels computed as negation of the set $T_i$.
\item \textbf{Non-Tumor Region} in week $i$ ($N^{\prime}_i$): Set of non-tumor voxels computed as negation of the set $T^{\prime}_i$.
\item \textbf{Unchanged Region} in week $i$ ($U_i$): The set of tumor voxels that is the intersection of $T_i$ and $T_{i+1}$ i.e, $U_i = T_i \cap T^{\prime}_{i+1}$.
\item \textbf{Reduced Region} in week $i$ ($R_i$): The set of tumor voxels that is defined as, $R_i = T_i \smallsetminus T^{\prime}_{i+1}$.
\item \textbf{Newly grown Region} in week $i$ ($G_i$): The set of tumor voxels that is defined as, $G_i = T^{\prime}_{i+1} \smallsetminus T_i$.
\end{itemize}

\begin{figure*}[!ht]
\centering
\subfloat[]{\includegraphics[width=0.33\textwidth,height=0.33\textwidth]{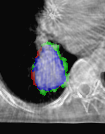}
\label{fig:Concost}}
\subfloat[]{\includegraphics[width=0.33\textwidth,height=0.33\textwidth]{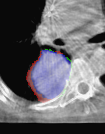}
\label{fig:RegCost}}
\subfloat[]{\includegraphics[width=0.33\textwidth,height=0.33\textwidth]{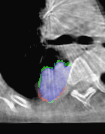}
\label{fig:SimCost}}
\caption{The images (a), (b), (c) show several slices of the CBCT images along with their delineations highlighted for weeks $i$ and $i+1$. The green region represents the reduced region ($R_i$), blue region represents the unchanged region ($U_i$), and red region represents the newly grown region ($G_i$), respectively.}
\label{fig:DispImage}
\end{figure*}

Applying the discussed model on real images, several slices are shown in Fig.~\ref{fig:DispImage}. We study the deformation fields in these regions.

\begin{table}
\centering
\caption{Bootstrap confidence intervals around the mean for the population of 29 patients.}
\label{tab:bootstrap_ci}
\begin{tabular}{|c|c|c|}
\hline
  & {\color[HTML]{3166FF} Confidence Interval} & {\color[HTML]{3166FF} Mean}\\ \hline
{\color[HTML]{3166FF} Reduced}     & [0.9957, 0.9982]   & 0.9969  \\ \hline
{\color[HTML]{3166FF} Newly Grown} & [1.0151,1.0175]  &  1.0174\\ \hline
{\color[HTML]{3166FF} Unchanged}   & [1.0388,1.0416]  &  1.0408 \\ \hline
{\color[HTML]{3166FF} Non-Tumor}   & [0.9881,0.9906] &  0.9895 \\ \hline
\end{tabular}
\end{table}

\subsection{Distribution of Jacobian}\label{stat_tests}
Statistical tests on the Jacobian are performed for the entire dataset on observing the distribution of J in regions of different categories as described in Section.~\ref{def_study}.
We assume normality of the dataset in all the regions due to the very large number of samples. The number of samples in $R$, $G$, $U$, and $N$ regions are approximately, $1.8$, $1.5$, $3.5$ and $169$ million, respectively.
The confidence intervals around the mean can be estimated as [$\bar{x} \pm 1.96 \frac{\sigma}{\surd{n}} $], for a confidence of 95\%. We also computed the confidence intervals for the different regions using bootstrap method. In bootstrapping, the samples are resampled to compute the confidence intervals. The bootstrap confidence intervals and the mean for a population of $29$ patients are shown in the Table.~\ref{tab:bootstrap_ci}. It can be observed that the intervals are very narrow.

\subsection{Two-sided t-test}\label{t-test}
The two-sided \emph{t}-test was performed under the null hypothesis that two independent distributions have identical expected values. The \emph{t}-test was done with an assumption that the two populations have unknown identical variances. The test gives a \emph{p}-value that explains the level of significance achieved. The \emph{p}-value gives the probability of achieving a result equal to or more extreme than what was observed, assuming the null hypothesis is true. A very low \emph{p}-value leads to the rejection of null hypothesis. The $p$-value was found to be close to zero for all the pair of regions. The \emph{t}-statistic reveals which of the two samples have higher or lower expected values.

Samples were collected from the different regions of the tumor and the test was conducted on the population of $29$ patients.
Let us denote the mean($\mu$) and standard deviation($\sigma$) in the regions with the respective subscripts, where $T$ corresponds to tumor regions, $N$ corresponds to non-tumor regions, $U$ corresponds to unchanged regions, $R$ corresponds to reduced regions, and $G$ corresponds to newly grown regions, respectively.

\begin{figure}[!ht]
\centering
\includegraphics[width=1.0\textwidth,height=0.8\textwidth]{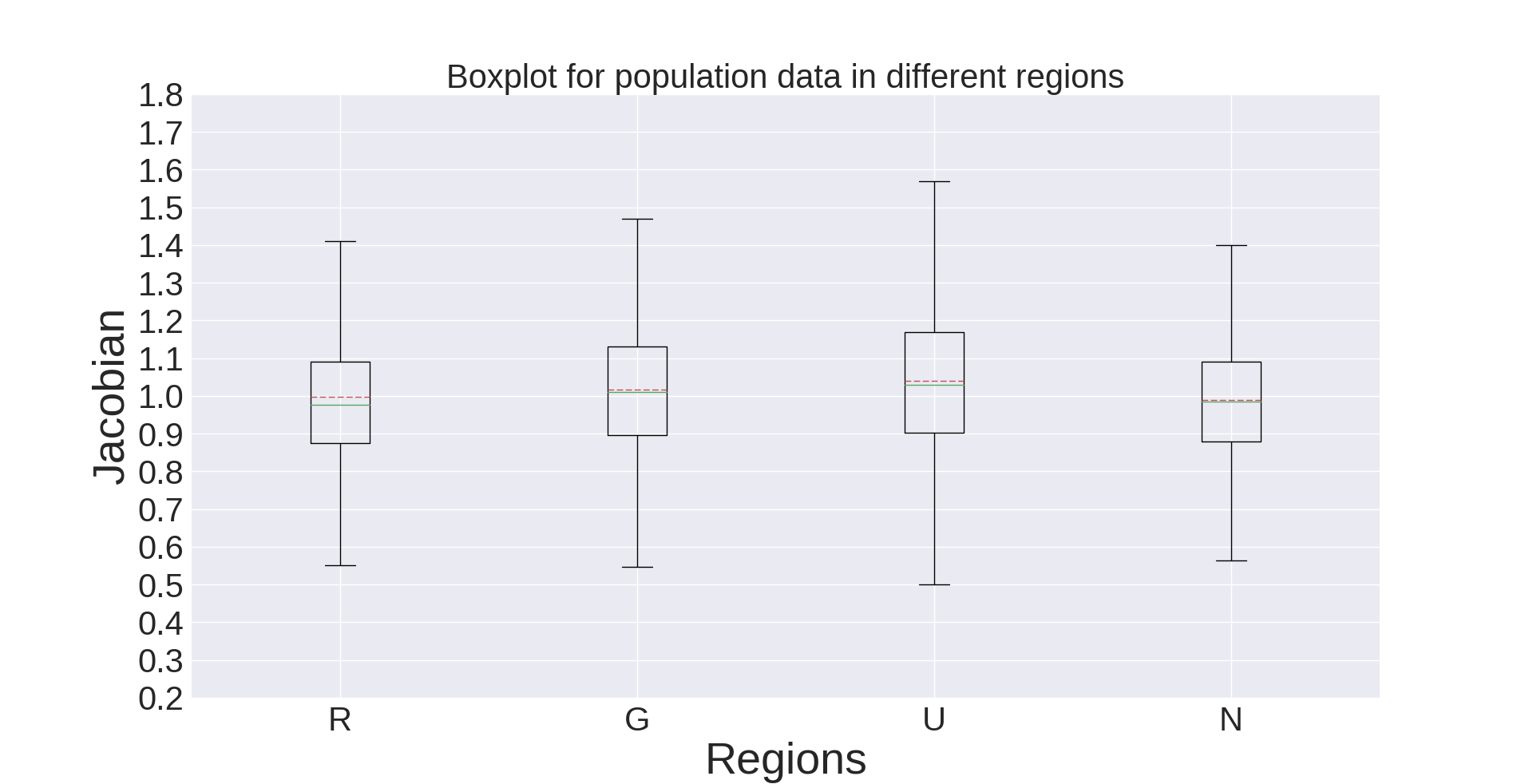}
\caption{Plot showing the mean(dotted line) and median(bold line) of the Jacobian for the population in different regions. The lower and upper quartile ranges along with the $98\%$ confidence intervals are also shown as whiskers in the box-plot.}
\label{fig:Boxplot_Population}
\end{figure}

The \emph{t}-statistic values are interpreted such that, when the alternative hypothesis is $\mu_X \geq \mu_Y$, then large positive values of t lead to rejection of the null hypothesis, where $X$ and $Y$ are defined as any of the four regions i.e, $X,Y \in {R,U,G,N} \mid X \neq Y$. Alternatively, when the \emph{t}-statistic values are large and negative, then $\mu_X \leq \mu_Y$. 
Fig.~\ref{fig:Boxplot_Population} shows the box-plot for the population in different regions.

The \emph{t}-statistic values are shown in Table.~\ref{tab:t_statistic_table}. Here, $X$ corresponds to the first column and $Y$ corresponds to the first row. It can be seen that $\mu_R \leq \mu_G$, $\mu_R \leq \mu_U$, and $\mu_G \leq \mu_U$, as the \emph{t}-statistic values are large and negative.
Similarly, $\mu_R \geq \mu_N$, $\mu_G \geq \mu_N$, and $\mu_U \geq \mu_N$, respectively, due to large positive values of the \emph{t}-statistic. 
By arranging the \emph{t}-statistic values on a real line, we observe that $\mu_N \leq \mu_R \leq \mu_G \leq \mu_U$ for the population. These results corroborate with the intuition also, as the $R$ region must be more aggressive than the $N$ region. $R$ region has recently reduced to non-tumor. Further, the $G$ region has recently manifested as tumor, hence we expect it to be aggressive; however, lower than the $U$ region which has currently not undergone any change. The above observation also is in conformation with the fact that $G$ and $U$ are relatively more aggressive than $R$ and $N$ regions. 

\subsection{Ordering of Jacobian for individual patients}\label{ordering-section}
Radiology Response Evaluation Criteria in Solid Tumors(RECIST) \cite{eisenhauer2009new} is used to assess the imaging of each patient and analyze response to radiation treatment.
Patients who have undergone treatment are categorized according to this criteria according to their response. The response of a patient is checked by radiation oncologists typically three months after completion of radiotherapy. A follow up CT scan is performed to check for progression. Based on the tumor size and metastasis, the patient is classified into one the categories as follows:
\begin{itemize}
  \item Complete Response (CR): Disappearance of all target lesions.
  \item Partial Response (PR): At least a 30\% decrease in the sum of the longest diameter(LD) of target lesions, taking as reference the baseline sum LD.
  \item Stable Disease (SD): Neither sufficient shrinkage to qualify for PR nor sufficient increase to qualify for PD, taking as reference the smallest sum LD since the treatment started (\% change between 30\% decrease and 20\% increase).
  \item Progressive Disease (PD): At least a 20\% increase in the sum of the LD of target lesions, taking as reference the smallest sum LD recorded since the treatment started or the appearance of one or more new lesions.
  \item Distant Progression (DP): The cancer is spreading from the original (primary) tumor in the prostate to lymph nodes or distant organs such as the bones, liver and lungs.
\end{itemize}

\begin{table}[!ht]
\centering
\caption{t-statistic obtained from a two sided t-test.}
        \label{tab:t_statistic_table}
        \begin{tabular}{|c|c|c|c|c|}
        \hline
         & {\color[HTML]{3166FF} R} & {\color[HTML]{3166FF} G} & {\color[HTML]{3166FF} U} & {\color[HTML]{3166FF} N} \\ \hline
        {\color[HTML]{3166FF}  R }     & -  & -139.21  & -289.84  & 66.24  \\ \hline
        {\color[HTML]{3166FF}  G } & 139.21 & -  & -123.50  & 249.77  \\ \hline
        {\color[HTML]{3166FF}  U }   & 289.84 & 123.50 & - & 540.96  \\ \hline
        {\color[HTML]{3166FF}  N }   & -66.24  & -249.77  & -540.96 & -   \\ \hline
        \end{tabular}
\end{table}

\subsubsection{Ordering hypothesis}\label{ordering-hypothesis}
We perform the \emph{t}-test for each of the patients to record the ordering of the means in different regions.
\begin{algorithm}[!ht]
\label{algo:ordering_algo}
\SetAlgoLined
\KwData{\emph{t}-test ordering result and Jacobian means in different regions for every patient}
\KwResult{Binary classification of patient response for each patient }
  \eIf{$\mu_R \leq 1.0$ and $\mu_R \leq \mu_U$ and $\mu_R \leq \mu_G$}{
   Classify patient as Partial Response(PR)\;
   }{
   Do not take a decision\;
  }
 \caption{Ordering hypothesis}
\end{algorithm}

\begin{figure*}[!ht]
\centering
\includegraphics[width=1.00\textwidth,height=0.8\textwidth]{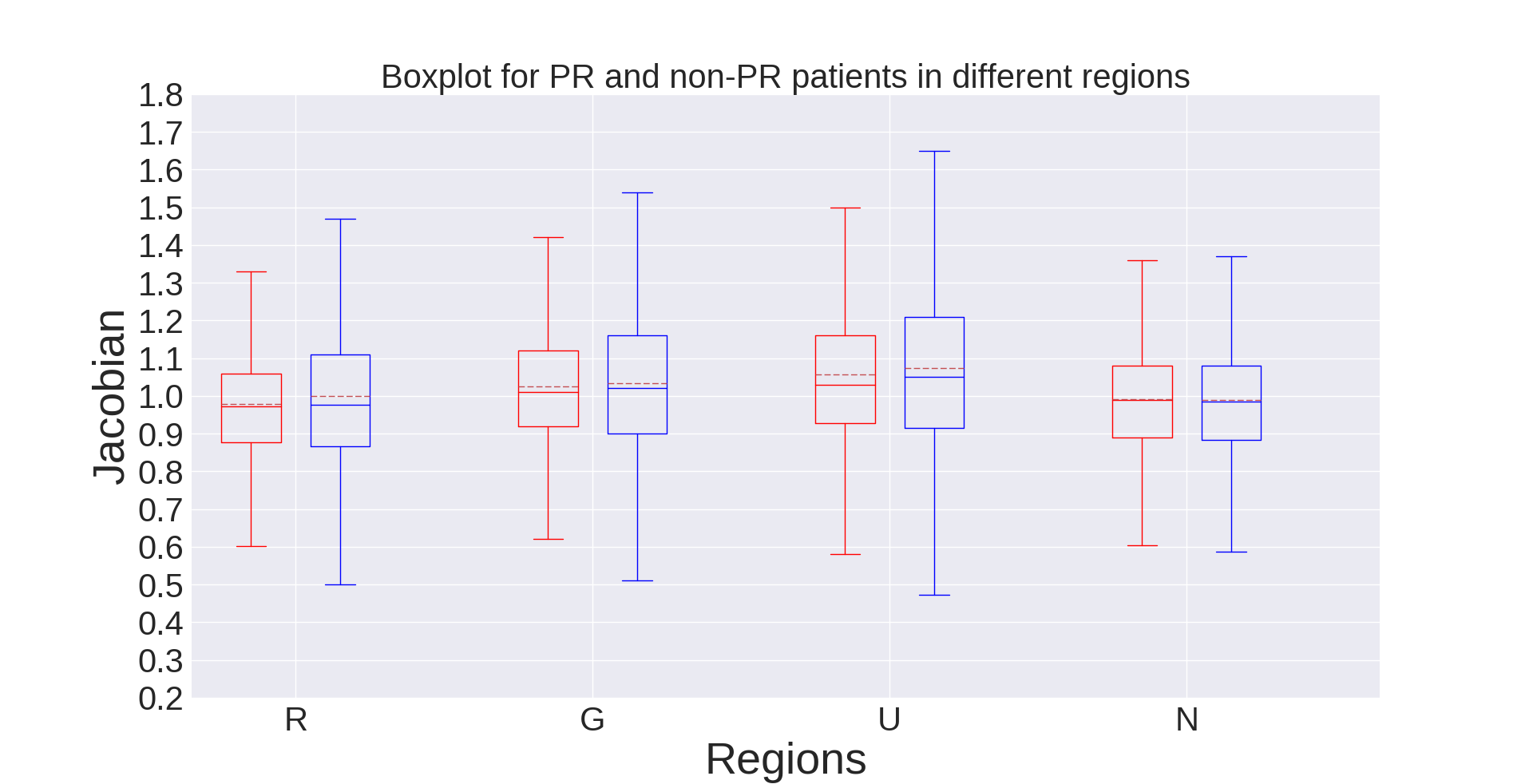}
\caption{Plot showing the mean(dotted line) and median(bold line) of the Jacobian for the entire dataset for PR and non-PR patients in different regions. The lower and upper quartile ranges along with the $98\%$ confidence intervals are also shown as whiskers in the box-plot.}
\label{fig:Boxplot_Hypothesis}
\end{figure*}

The hypothesis to classify each patient is explained in Algorithm.~\ref{algo:ordering_algo}.
The aim is to correctly classify partial response(PR) patients using the hypothesis.
It is to be noted that, by definition, a CR qualifies as a PR but a PR does not qualify as a CR.
Therefore, in our analysis PR category of classification refers to patients with response as either PR or CR, while the non-PR category refers to patients with response neither as PR nor as CR.
We perform the \emph{t}-test for each patient in the population and test set.
The ordering from \emph{t}-test is modified in the hypothesis for classifying patients into PR and non-PR categories. It is designed based on the following logic: By definition, we know that the expansion in $G$ and $U$ should be greater than that of $R$, because of more tumorous activity in these regions. Therefore, we hypothesize that $\mu{R} \leq \{\mu_{G},\mu_{U}\}$. This hypothesis does not consider the ordering between $G$ and $U$. Additionally we add another condition based on the tumor activity in the $R$ region. Intuitively, since $R$ region should result in net reduction of the volume, the mean of Jacobian $\mu_R$ should be less than $1.0$.
It can be observed from Fig.~\ref{fig:Boxplot_Hypothesis}, that the mean of patients for PR is less than 1.0, while that of NPR is close to 1.0.

\begin{figure*}[!ht]
\centering
\includegraphics{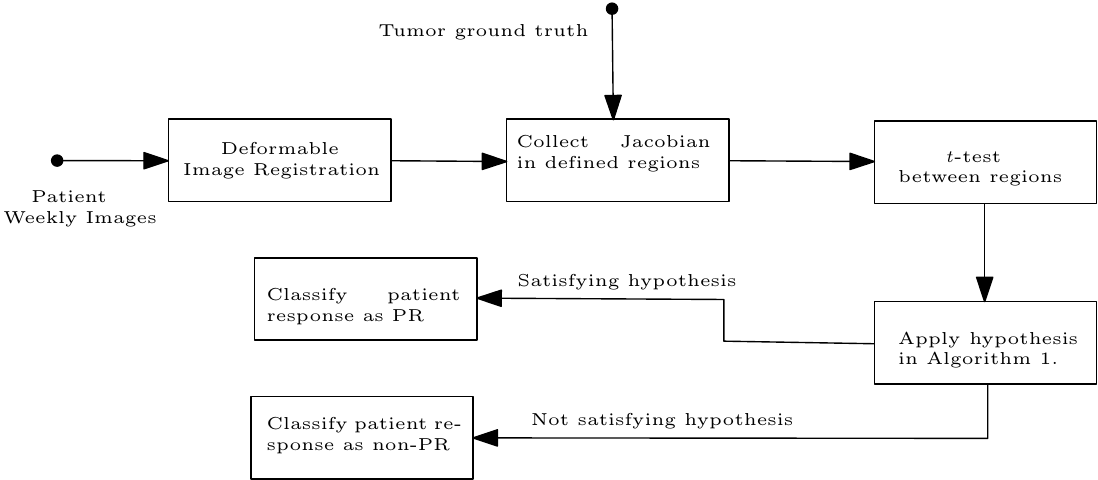}
\caption{The overall picture of classifying each patient in the population and test data is shown here as a block diagram.}
\label{fig:BlockDiagram}
\end{figure*}

Fig.~\ref{fig:BlockDiagram} shows the block diagram of the classification process.
Our data consists of the radiation response of each patient in terms of the RECIST criteria. Table.~\ref{tab:ordering_full} shows the corresponding radiation response(RX\_Response) for each patient. Patients whose response to radiation could not be determined by doctors are labeled as ``NA". 

Based on this hypothesis, we label each patient in the ``Classification" column of Table.~\ref{tab:ordering_full} in ~\ref{ordering-classification}. We compare the predicted and the actual response(RX\_Response) to summarize into Table.~\ref{tab:summary_ordering_full} and Table.~\ref{tab:summary_ordering_3w}, while excluding the patients whose response was ``NA".
Overall, we find that $12$ out of $21$ patients are correctly classified as PR, and $13$ out of $17$ are correctly classified as non-PR patients.

\begin{table}[!ht]
\centering
\caption{Contingency matrix for the full data.}
\label{tab:summary_ordering_full}
\begin{tabular}{|c|c|c|}
\hline
  & {\color[HTML]{3166FF} Partial Response} & {\color[HTML]{3166FF} Not Partial Response}\\ \hline
{\color[HTML]{3166FF} Hypothesis satisfied}     & 12  &  4   \\ \hline
{\color[HTML]{3166FF} Hypothesis Not satisfied} & 9   &  13  \\ \hline
\end{tabular}
\end{table}

\begin{table}[!ht]
\centering
\caption{Contingency matrix for first three weeks data.}
\label{tab:summary_ordering_3w}
\begin{tabular}{|c|c|c|}
\hline
  & {\color[HTML]{3166FF} Partial Response} & {\color[HTML]{3166FF} Not Partial Response}\\ \hline
{\color[HTML]{3166FF} Hypothesis satisfied}     & 11  &  3   \\ \hline
{\color[HTML]{3166FF} Hypothesis Not satisfied} & 10   &  14  \\ \hline
\end{tabular}
\end{table}

Fisher's exact test for association between PR and the hypothesis was performed on the contingency table Table.~\ref{tab:summary_ordering_full} and Table.~\ref{tab:summary_ordering_3w}, which yields an odds ratio(OR) and a \emph{p}-value, to measure the association between two categories PR and non-PR. The null hypothesis is that the occurrence of PR and non-PR are equally likely. The odds ratio(OR) measures the ratio of odds of occurrence of an event to the odds of an event not occurring. The contingency table Table.~\ref{tab:summary_ordering_full} corresponding to complete six weeks data yielded an OR and \emph{p}-value of 4.33 and 0.051, respectively.

Using only the first three weeks data, we find that $11$ out of $21$ patients are correctly classified as PR, and $14$ out of $17$ are correctly classified as non-PR patients. From Table.~\ref{tab:summary_ordering_3w}, for the first three weeks data, the obtained result for OR and \emph{p}-value were 5.13 and 0.043, respectively. Assuming a critical \emph{p}-value of 0.10 as threshold, we can reject the null hypothesis that there is similarity in occurrence between PR and non-PR categories. The OR of 5.13 represents that it is 5.13 times more likely that PR happens when the hypothesis is satisfied than non-PR when the hypothesis is satisfied. This analysis suggests that the hypothesis is satisfied for patients with response as PR during the first three weeks of the treatment also. This early prediction can help doctors take necessary actions for improving the response of the patient to treatment. 

The accuracy, precision and recall were found to nearly the same on the three weeks data as shown in Table.~\ref{tab:summary_overall}. The level of significance is close to being significant considering the full six weeks and first three weeks of the dataset.

\begin{table}[!ht]
\centering
\caption{Summary of results for full and three week data.}
\label{tab:summary_overall}
\begin{tabular}{|c|c|c|}
\hline
  & {\color[HTML]{3166FF} Full} & {\color[HTML]{3166FF} Three Weeks}\\ \hline
{\color[HTML]{3166FF} Accuracy}     & 65.7  &  65.7   \\ \hline
{\color[HTML]{3166FF} Recall} & 60.0   &  52.4  \\ \hline
{\color[HTML]{3166FF} Precision} & 75.0   &  78.6  \\ \hline
{\color[HTML]{3166FF} Odds Ratio} & 4.33   &  5.13  \\ \hline
{\color[HTML]{3166FF} Level of significance} & 0.051   &  0.043  \\ \hline
\end{tabular}
\end{table}

\section{Conclusion}
In this work we discussed analysis of Jacobian obtained from the deformation fields of registering spatiotemporal data of CBCT images for patients who underwent radiotherapy. Using clinical radiation oncologists delineation over the first three weeks of treatment duration, we analyzed the behavior of Jacobian in four regions of each delineated tumor. We observed very narrow variation of the Jacobian in the population in each of the regions. Two-sided t-test was performed to identify the ordering of mean Jacobian for the population in the four regions as described above. Based on the ordering obtained from the population, we proposed a hypothesis for classification of each patient within the population and test data into PR and non-PR classes. This hypothesis was used to segregate patients into those satisfying the hypothesis and not satisfying the hypothesis. We observed that patients satisfying the proposed hypothesis had better RECIST response to radiation treatment. Significant association between the proposed hypothesis and better response was confirmed by using Fisher's test. 
This early prediction has significant impact in providing radiologists necessary feedback regarding the response of patient to the treatment. Early prediction before the stipulated radiation course of 6 weeks gives enough time for doctors to take alternate actions like increased radiation dose, or alternative treatment approaches like chemotherapy, surgery. Using this result, the images of new patients who have undergone treatment can be processed using the discussed work flow, to qualitatively predict the response of a treatment. Further analysis can be performed to predict quantitative measures that can indicate prognosis for the patient.

\section*{Acknowledgment}
This work is carried out under the MHRD sponsored project entitled as
``Predicting Cancer Treatment outcomes of lung and colo-rectal cancer by modeling and analysis of anatomic and metabolic images". We would additionally like to thank, Partha Sen and Gaurav Goswami for helping us procure the data and validating the image registrations.

\section*{References}

\bibliography{references}

\appendix
\section{Details of the dataset}\label{ordering-classification}
\begin{table}[h]
\centering
\caption{Details of each patient output to the hypothesis and the corresponding actual response.}
\label{tab:ordering_full}
\resizebox{1.0\columnwidth}{!}{%
\begin{tabular}{|c|c|c|c|c|c|c|c|}
\hline
  {\color[HTML]{3166FF} Patient ID} & {\color[HTML]{3166FF} Classification(Full) }  & {\color[HTML]{3166FF} Classification(3 Weeks)} & {\color[HTML]{3166FF} RX Response} & {\color[HTML]{3166FF} Patient ID} & {\color[HTML]{3166FF} Classification(Full) }  & {\color[HTML]{3166FF} Classification(3 Weeks)} & {\color[HTML]{3166FF} RX Response} \\ \hline
{ 1 } & Y & Y & PR & { 24 } & N & N & PR \\ \hline
{ 2 } & Y & Y & NA & { 25 } & N & N & NA \\ \hline
{ 3 } & Y & Y & PR & { 26 } & N & N & PR \\ \hline
{ 4 } & N & N & PR & { 27 } & N & N & PR \\ \hline
{ 5 } & Y & Y & NA & { 28 } & Y & Y & PR \\ \hline
{ 6 } & Y & Y & PR & { 29 } & Y & N & NA \\ \hline
{ 7 } & N & N & NA & { 30 } & N & N & PD \\ \hline
{ 8 } & Y & Y & PR & { 31 } & Y & Y & PD \\ \hline
{ 9 } & N & N & PD & { 32 } & N & N & PD \\ \hline
{ 10 } & N & N & SD & { 33 } & Y & Y & SD \\ \hline
{ 11 } & N & N & SD & { 34 } & Y & Y & SD \\ \hline
{ 12 } & N & N & NA & { 35 } & N & N & SD \\ \hline
{ 13 } & Y & N & PD & { 36 } & Y & Y & PR \\ \hline
{ 14 } & N & N & PR & { 37 } & Y & Y & PR \\ \hline
{ 15 } & N & N & PR & { 38 } & N & N & PD \\ \hline
{ 16 } & N & N & SD & { 39 } & N & N & PR \\ \hline
{ 17 } & Y & Y & PR & { 40 } & N & N & SD \\ \hline
{ 18 } & Y & Y & NA & { 41 } & N & N & SD \\ \hline
{ 19 } & N & N & SD & { 42 } & N & N & SD \\ \hline
{ 20 } & N & N & PR & { 43 } & Y & N & CR \\ \hline
{ 21 } & Y & Y & PR & { 44 } & Y & Y & CR \\ \hline
{ 22 } & N & N & PD & { 45 } & N & N & CR \\ \hline
{ 23 } & Y & Y & PR & \\ \hline
\end{tabular}
}
\end{table}

\end{document}